\documentclass[preprint,showpacs,preprintnumbers,amsmath,amssymb]{revtex4-1}

\usepackage{graphicx}
\usepackage{dcolumn}
\usepackage{bm}
\usepackage{amsmath}
\usepackage{subfigure,empheq}
\usepackage{hyperref}
\usepackage{natbib}

\begin{document}
\title{Synchronization, phase slips and coherent structures in area-preserving maps}
\author{Swetamber Das}
\email{swetdas@physics.iitm.ac.in}
\affiliation{Department of Physics, Indian Institute of Technology Madras, Chennai, 600036, India.}
\author{Sasibhusan Mahata}
\email{sasibhusan@physics.iitm.ac.in}
\affiliation{Department of Physics, Indian Institute of Technology Madras, Chennai, 600036, India.}
\author{Neelima Gupte}
\email{gupte@physics.iitm.ac.in}
\affiliation{Department of Physics, Indian Institute of Technology Madras, Chennai, 600036, India.}
\begin{abstract}

{\footnotesize The problem of synchronization of coupled Hamiltonian systems exhibits interesting features due to the non-uniform or mixed nature (regular and chaotic) of the phase space. We study these features by investigating  the  synchronization of unidirectionally coupled area-preserving maps coupled by the Pecora-Carroll method.  We find that coupled standard maps show complete synchronization for values of the nonlinearity parameter at which regular structures are still present in phase space. The distribution of synchronization times has a power law tail indicating long synchronization times for at least some of the synchronizing trajectories.  With the introduction of coherent structures using parameter perturbation in the system, this distribution crosses over to exponential behavior, indicating shorter synchronization times, and the number of initial conditions which synchronize increases significantly, indicating an enhancement in the basin of synchronization. On the other hand, coupled blinking vortex maps display both phase synchronization and phase slips, depending on the location of the initial conditions. We discuss the implication of our results.}
\end{abstract}
\maketitle

\section{Introduction}
Hamiltonian and area-preserving systems show several interesting features such as the existence of anomalous kinetics, L\'{e}vy processes and L\'{e}vy flights \cite{Klafter1994}, power law contributions to recurrence and other statistics, and the existence of dynamical traps \cite{Zaslavsky2002a,Zaslavsky2002b}, due to the mixed nature of their phase space. Our interest here in the phenomenon of synchronization which shows different features in the case of Hamiltonian systems, compared to those found in the case of dissipative systems. In this paper, we study the synchronization behavior of area-preserving maps, as models of Hamiltonian systems, where the maps are synchronized using the coupling scheme of Pecora and Carroll \cite{Pecora1990}. The synchronization of dissipative systems using this method is well studied, both in theoretical and experimental contexts \cite{Pecora1991, Roy1994, Abarbanel1995, Kocarev1996, Brown1997, Pecora1997,Ulrichs2009, Lopes2014}. However,  there are very few studies on synchronization in Hamiltonian or area-preserving systems in the literature. 

We note that two dynamical systems will be taken  to be completely \textit{synchronized} if the difference between any of their quantifiable state properties converges to zero as $t \rightarrow \infty$. Phase synchronization in the system, on the other hand, can be demonstrated by analyzing their phases via Hilbert transform of their state variables~\cite{Pikovsky2001,Boccaletti2002}. More generalized definitions of synchronization can be found in the numerous reviews on the subject \cite{Pikovsky2001, Boccaletti2002, Pecora2015}.

In this paper, we study the effects of the mixed phase space on the synchronization properties of Hamiltonian systems, as represented by area-preserving maps. We demonstrate the procedure using the standard map \cite{Chirikov1979}, the prototypical area-preserving map, and then look at behavior in the blinking vortex (BV) map, a  map model for a stirred tank reactor \cite{Aref1984} as our examples. We  find that the distribution of synchronization times, for coupled standard maps, has a power law tail, for randomly chosen combinations of initial conditions which lead to regular/chaotic trajectories. The introduction of a coherent structure in the phase space leads to a huge reduction in synchronization times, and a corresponding exponential distribution for the synchronization times, as well as a substantial increase in the number of initial conditions which synchronize. In the case of the BV map, one set of initial conditions synchronizes, whereas another leads to phase slips; reminiscent of phase slips that are seen in experiments on thermally coupled rotating fluid annuli \cite{Pita2010} which model atmospheric convection. We comment on the significance of our results.  

This paper is organized as follows.  In sec.~\ref{The coupled  standard map}, we describe the simple unidirectional coupling scheme used to couple the area-preserving maps under investigation and compute the Lyapunov exponents from the master stability function for the coupled system. In sec.~\ref{The effect of a coherent structure}, we will introduce a \text{coherent structure} in the coupled system and study its effect on synchronization and synchronization times. Next, sec.~\ref{Coupled Blinking Vortex map} explores  synchronization in the blinking vortex map which displays phase slips as well as phase synchronization  effects. Our results and their implications are discussed  in sec.~\ref{sec:conclusion} 

\section{The coupled  standard map}
\label{The coupled  standard map}

The standard map is considered to be the prototypical example of a two-dimensional area-preserving map, and is given by:
\begin{empheq}[right=\empheqrbrace \mod 2\pi]{align}
\hspace{2.2cm}P_{n+1} &= P_n + K\sin Q_n \nonumber\\
Q_{n+1} &= P_{n+1} + Q_n 
\end{empheq}
Here the subscript $n$ denotes the discrete time and $K$ is the nonlinearity parameter. These equations typically describe the evolution of two canonical variables $P$ and $Q$ which correspond to the momentum and co-ordinate in the Poincar\'{e} section of a kicked rotator, a system which represents the behavior of a variety of mechanical systems, as well as the behavior of accelerator systems.  

We now synchronize two standard maps, using the Pecora-Carroll scheme of synchronization using drive-response coupling \cite{Pecora1990}. 
This system was first devised to synchronize dissipative chaotic dynamical systems, and first demonstrated that even chaotic trajectories can be effectively synchronized. Under this unidirectional coupling scheme, we duplicate the given map and couple the original and the duplicated map in a drive - response configuration. This means that the drive map evolves freely but the evolution of the response map is dependent on the drive. In this case, the $P$ value of the response system is set to the $P$ value of the drive system, at each iterate.  The initial values of $Q$ in the drive and response maps are chosen arbitrarily, whereas the $P$ values are identical. The system is said to reach complete synchronization when both the $P$ and $Q$ values of the drive and response systems evolve to identical values.  We analyze the synchronization of this system using the master stability function~\cite{Pecora1998}. 
\subsection{Master stability function}
\label{Master stability function}
A general drive-response system coupled unidirectionally may be described by the following set of equations

\begin{align}
\hspace{1cm} \frac{d\mathbf{X}_d}{dt} = \mathbf{F}(\mathbf{X}_d), \nonumber  &&
\frac{d\mathbf{X_r}}{dt} = \mathbf{F}(\mathbf{X}_r) + \alpha E(\mathbf{X}_d-\mathbf{X}_r)
\end{align}

Here $\mathbf{X}_d$ and $\mathbf{X}_r$ are drive and response variables; the matrix $E$ determines the linear combination of the $\mathbf{X}$ used in the difference and $\alpha$ is the coupling strength. For the map case, we have have the following form
\begin{align}
\hspace{1cm}  \mathbf{X}^d_{n+1} = \mathbf{F}(\mathbf{X}^d_{n}), \nonumber  &&
\mathbf{X}^r_{n+1} = \mathbf{F}(\mathbf{X}^r_{n}) + \alpha E(\mathbf{X}^d_{n}-\mathbf{X}^r_n)
\end{align}

To find the stability of the synchronous state, we first express  the system in terms of  $P^\perp = P^d - P^r$ and $Q^\perp = Q^d - Q^r$, as follows
\begin{empheq}{align}
\label{equ:trans}
\hspace{.3cm} P^\perp_{n+1} &= (1-\alpha)P^\perp_n+ K \sin(Q^d_n) - K \sin(Q^r_n) \nonumber \\
Q^\perp_{n+1} &= (1-\alpha)P^\perp_n+ Q^\perp_n + K \sin(Q^d_n) - K \sin(Q^r_n)  
\end{empheq}

Here, the superscripts $d$ and $r$ indicates the $P$ and $Q$ values corresponding to drive and response maps. The associated largest LE is the master stability function of the system, and is given by:
\begin{equation}
\label{equ:LE}
\hspace{2cm}  \lambda   = \lim_{n \rightarrow \infty} \frac{1}{n}\sum^{n-1}_{i=0}\ln\big|1+K\cos(Q^d_n)\big|
\end{equation}

\begin{figure}[h!]
\centering
\includegraphics[height=7.5cm,width = 10cm]{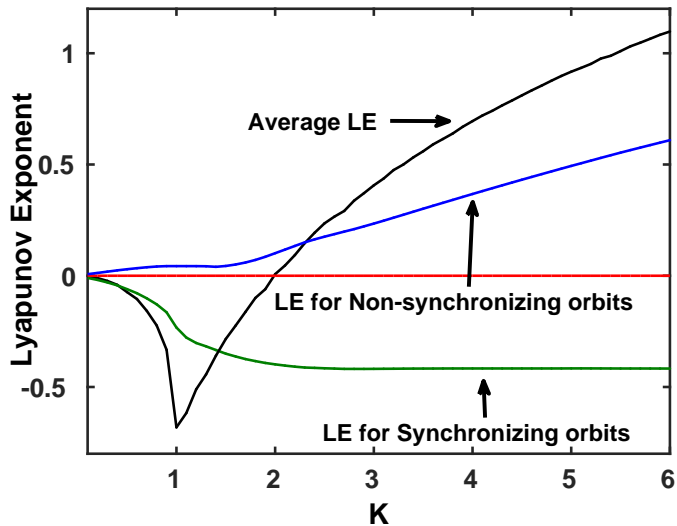} 
\caption{\label{fig:LE} \footnotesize  The variation of the Lyapunov exponent as a function of K. The LE's have been calculated and averaged over 5 000 initial conditions randomly chosen from the interval $[0,2\pi]$ with 5 000 iterations in each case. It is seen that the average LE value crosses the zero line at $K = 2$ and remain positive afterwards. The three curves show the variations of LE for synchronizing and non-synchronizing orbits along with the LE for all the orbits as indicated.}
\end{figure}

\begin{figure}[t!]
\centering
\includegraphics[height=7cm,width = 10cm]{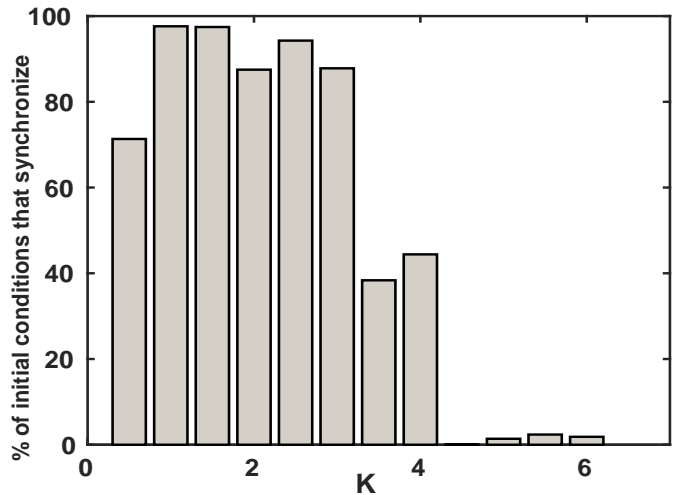}
\caption{\label{fig:Sync_hist} \footnotesize The percentage of the initial conditions that converge to synchronization for $K = (0.5,1,1.5,...,6)$. We have chosen 10 000 initial conditions randomly chosen over the interval $[0,2\pi]$ with 20 000 iterations in each case. The synchronization accuracy used is $10^{-5}$.}
\end{figure}

The variation of the LE associated with the $Q$ direction is shown in Fig.~\ref{fig:LE} in which the LE is calculated and averaged over 5 000 randomly chosen initial conditions for different values of $K$. There is no LE is in the $P$-direction as $P_d$ and $P_r$ are identical under the Pecora-Carroll coupling scheme. In Fig.~\ref{fig:LE}, we observe that the LE is negative upto $K = 2$, and crosses to positive values thereafter. However, due to the mixed nature of the phase space,  synchronization can occur even for the values of $K$ where the average LE is not negative depending on the initial conditions used. The plot in Fig.~\ref{fig:LE} also shows the LE exponents for synchronizing initial conditions and non-synchronizing initial conditions separately. The numerical accuracy used to determine synchronization in our computations is $10^{-5}$ i.e. synchronization is assumed to be reached if $|Q^d-Q^r| \leq 10^{-5}$ for all subsequent iterations. We plot the percentage of initial conditions which evolve to the synchronized state in Fig.~\ref{fig:Sync_hist}. Trajectories that have not synchronized after 5 000 iterations are counted as non-synchronized.  It is clear that the LE is negative for synchronizing initial conditions and positive for non-synchronizing ones as it should be.

We find that more than 70\% of initial conditions synchronize for $K\leq 3$ but for higher values of $K$, the synchronizing fraction  drops to almost 2\% of the whole. We also note that a sizeable fraction of initial conditions lead to synchronization at $K = 2.5$ and $K = 3$ where the average value of the LE is positive (compare with Fig. \ref{fig:LE}). This happens due to the fact that the LE-s for non-synchronizing orbits having non-negative values are larger in magnitude, and are also weighted by the fraction of nonsynchronizing initial conditions, which contributes to the increase in the average. Therefore, we obtain synchronization for some initial conditions even if the {\it average} LE is non-negative. In addition, we do not find  any synchronization at $K=4.5$; consistent with the corresponding LE. It is clear that synchronization in this system shows a strong dependence on the initial conditions. A detailed study of this has been reported in our earlier work~\cite{Mahata2016}.

\begin{figure}[t]
\includegraphics[height=7.5cm,width = 10cm]{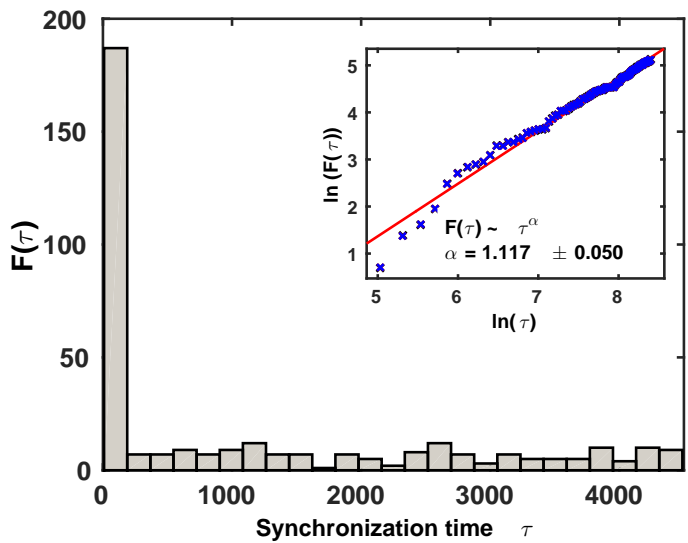} 
 \caption{\label{fig:Long_tail} \footnotesize The distribution of synchronization times for the coupled standard map at K = 6. The distribution has 25 bins of synchronization times for 50 000 different initial conditions randomly chosen from the uniform distribution on the interval $[0,1]$. \textit{Inset} -- The long tail in the distribution shows a short power law regime as seen in the log-log plot of the cumulative distribution. The exponent $\alpha$  for the fit shown takes the value $1.117\pm 0.050$.} 
\end{figure}

\subsection{Synchronization times}
\label{sync_times}

Synchronization time, by definition, is the number of iterations required to reach complete synchronization in a coupled system under a given numerical accuracy. We note that there exists a variety of factors on which synchronization depends~\cite{Mahata2016}. These include (a) the location of regular orbits (sticky regions), (b) the values of the nonlinearity parameter $K$, (c) initial conditions, (d) the number of iterates, and (e) the numerical accuracy assumed for synchronization. 

 It is observed that with the increase of the nonlinearity parameter $K$, the number of initial conditions that show synchronization decreases. For instance, at $K = 6$, only about $0.7\%$ of the chosen initial conditions lead to synchronized behavior. The synchronization time distribution for all the pairs here collectively shows a very long tail as in Fig.~\ref{fig:Long_tail} which is plotted for 50 000 different initial conditions randomly chosen from the uniform distribution on the interval [0,1]. Interestingly, the tail in the distribution shows a short power law scaling (the inset in Fig.~\ref{fig:Long_tail}) with the exponent of the cumulative distribution being given by $1.117\pm 0.030$. The tail of the distribution corresponds to long synchronization times, and the power law decay indicates that in addition to the fact that a very small number of initial conditions synchronize here, a significant fraction of these exhibit a very slow rate of convergence. A detailed investigation of the effect of combinations of initial conditions (regular/chaotic) for drive and response maps is given in Ref.~\cite{Mahata2016}.

We note that the phase space structure seen here is fairly homogeneous at this high value of the nonlinearity parameter $K$, with most of the phase space being accessed by chaotic trajectories, and the exponents settle down more stable values. On the other hand, for lower values of $K$, the phase space is inhomogeneous as many periodic orbits exists in the space. This leads to a greater dependence of the power - law exponent on the number of iterates.  It is interesting to see whether the introduction of a large inhomogeneity in the phase space, can contribute to an increase in the basin of synchronization, and to a reduction in the synchronization times. In the next section, we will  generate a large inhomogeneity in the phase space by creating a coherent structure in the system through a parameter perturbation method. 

\section{The effect of a coherent structure on synchronization}
\label{The effect of a coherent structure}

 Coherent structures are regular and localized structures in the phase space of chaotic and turbulent dynamical systems. These structures are stable and not affected by the chaotic nature of the system. A coherent structure can be generated in the area-preserving standard map using the method of parameter perturbation \cite{Gupte2007}. The perturbation is applied in the local neighborhood of the periodic points of the map. The procedure keep the area-preserving nature of the map intact. We describe the method briefly by taking the following form of the standard map
\begin{empheq}[right=\empheqrbrace \mod 1]{align}
\hspace{2.2cm}P_{n+1} &= P_n + \frac{K}{2\pi}\sin(2\pi Q_n) \nonumber\\
Q_{n+1} &= P_{n+1} + Q_n 
\end{empheq}

\begin{figure}[b!]
\includegraphics[height=7.5cm,width = 10cm]{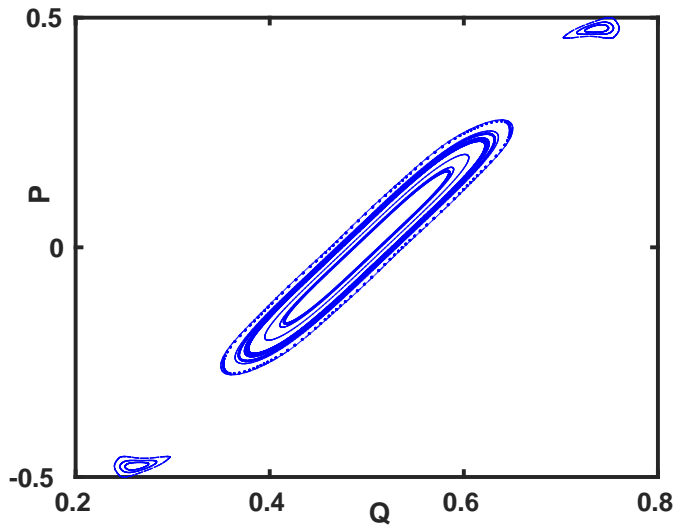} 
\caption{\label{fig:Coherent} \footnotesize The coherent structure in the phase space of the standard map for $K = 6$ , $\delta = 0.3$ and $\epsilon = 2$ about the fixed point $(P_f,Q_f)=(0.0,0.5)$. See the text for details of the method of parameter perturbation.  We evolve 25  initial conditions chosen randomly from the uniform distribution on the interval [0,1] for 4500 iterations, after 500 iterates, which are discarded as transients. Synchronization is seen in the phase space of the drive and response maps in the presence of the coherent structure.}
\end{figure}

Here, $-0.5 \leq P_n \leq 0.5, 0\leq Q_n \leq 1$ and  $n$ denotes the discrete time and $K$ is the nonlinearity parameter as seen previously. In the form above, the standard map is known to have a hyperbolic fixed point at $(0,0.5)$. We perturb the parameter $K$ to $K-\epsilon$ if $|P - P_f| < \delta$, $|Q-Q_f| < \delta$. For $P$ and $Q$ outside this $\delta$-strip, $K$ does not change.  It may be noted that the Jacobian $J$ of the original standard map is given by 
\[ \hspace*{2cm} J = \left( \begin{array}{cc}
1 & 1 + K\cos(2\pi Q_n) \\
1 & K\cos(2\pi Q_n) \end{array} \right).\] 
The determinant of this matrix is 1 as the dynamics is area-preserving. This remains unity for the perturbed map also in which $K$ is replaced by $K-\epsilon$ \cite{Sharma1997}. The coherent structure thus obtained for $K = 6$ , $\delta = 0.3$ and $\epsilon = 2$ about the fixed point $(P_f,Q_f)=(0.0,0.5)$  is shown in Fig~\ref{fig:Coherent}.

\begin{figure}[h!]
\begin{center}
 \includegraphics[height=7.5cm,width = 10cm]{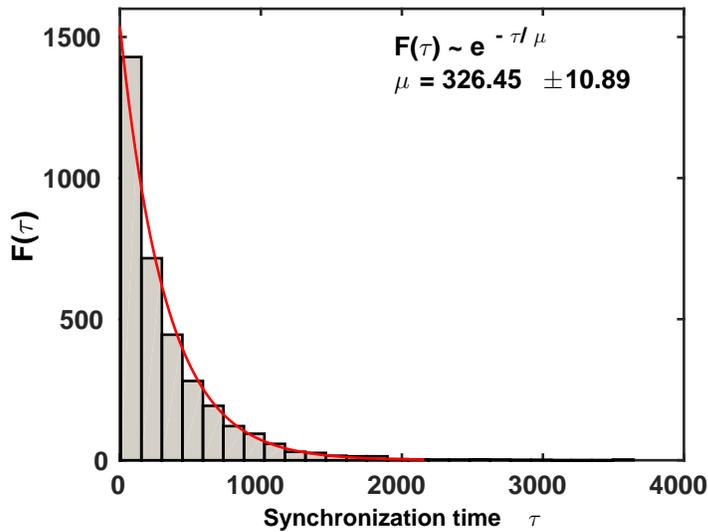}
 \caption{\label{fig:Expo} \footnotesize The distribution for synchronization times of the coupled standard map system shows an exponential distribution as shown in the figure, in the presence of a perturbation leading to a coherent structure created using the parameter perturbation technique. The perturbation is applied in the neighborhood of the fixed point $(P_f,Q_f)=(0.0,0.5)$ with $K = 6$ , $\delta = 0.3$ and $\epsilon = 2$. The distribution has 25 bins for synchronization times for 50000 different initial conditions randomly chosen from the uniform distribution on the interval [0,1]. The exponent of the distribution is found to be $\mu=326.45 \pm 10.89$. The mean synchronization time for this case is $326.44$.}
\end{center}
\end{figure}

We now examine the synchronization of the perturbed maps above, when the maps  are coupled using the Pecora-Carroll method. With the introduction of a coherent structure in the coupled system at $K = 6$ with the perturbation parameters as above, the response map also develops an identical structure in the phase space due to synchronization. Here, the perturbation is applied to both the drive and the response maps. We find that synchronization times reduce remarkably after the introduction of the coherent structure.  This indicates that the parameter perturbation forces the trajectories of the drive and response maps to stick around the coherent structure and therefore leads to considerably shorter synchronization times. In addition to this, we observe that the percentage of initial conditions which show synchronization increases to about $7\%$, which is an increase by a factor of $10$ as compared with the fraction of initial conditions which synchronize seen in the case of the unperturbed map. Thus, the basin of synchronization is greatly enhanced. In quantitative terms, we find that the distribution of synchronization times is now exponential as shown in Fig~\ref{fig:Expo} with the exponent $\mu=326.45 \pm 10.89$ for 50000 different initial conditions randomly chosen from the uniform distribution on the interval [0,1]. Therefore, the heavily long tailed distribution seen in the unperturbed map (Fig~\ref{fig:Long_tail}) turns into an exponential distribution in the presence of the coherent structure induced by the perturbation (Fig~\ref{fig:Expo}). 

If the phase space of the unperturbed map is predominantly chaotic (as at $K=6$), the creation of a coherent structure leads to a much larger enhancement in the fraction of synchronizing initial conditions (a factor of $10$ at $K=6.0$), than for the case where the mixed phase space has more regular regions in the first place, as at $K = 1.5$ where regular regions occupy about half of the phase space. Here, the creation of coherent structures enhances the number of synchronizing initial conditions by only about $20 \%$. The size of the newly created structure also has a role to play in this enhancement, as  larger coherent structures lead to shorter synchronization times. Therefore, this procedure provides a method to drastically reduce synchronization times and extend the basin of synchronization in coupled standard maps. We now discuss coupled blinking vortex maps which display the phenomenon of phase slips, in addition to phase synchronization. Such phase slips were not observed in coupled standard maps.  

\section{The coupled Blinking Vortex map:  Phase synchronization and phase slips}
\label{Coupled Blinking Vortex map}

\begin{figure}[t!]
 \includegraphics[height=7.5cm,width = 10cm]{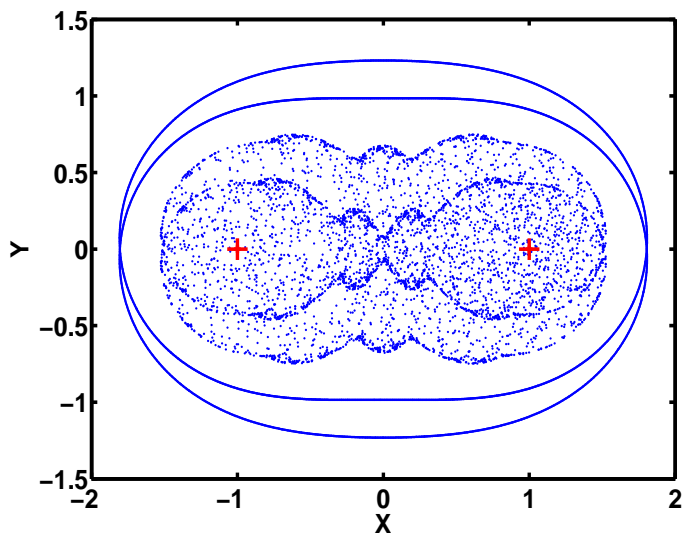}
 \caption{\label{fig:BV} \footnotesize The phase space of the BV map at $\mu=0.55$ for 25 randomly chosen initial conditions from the uniform distribution  $\in [-2,2]$ with 10000 iterations (500 of which are discarded as transients). The `+' sign (in red) indicates the position of the vortices.}
\end{figure}

The Blinking Vortex (BV) map studied in this section, is the simplest idealization of a stirred tank where  {\it chaotic advection} is seen \cite{Aref1984} in a two dimensional context. The system contains two agitators for stirring modeled by point vortices held at fixed positions. The two point vortices are sources of unsteady potential flow and operate alternately. This unsteady flow generates chaotic advection of the passive scalars present around the vortices which leads to efficient stirring. This two - dimensional dynamical system is known to reproduce the characteristics of realistic fluid flow.  The map equations for this system are 
\begin{empheq}{align}
\hspace{.8cm}y_{n+1} &= (x_n-\xi_i)\sin\left(\frac{\mu}{r^2}\right) + y_n\cos\left(\frac{\mu}{r^2}\right) \nonumber \\
x_{n+1} &= \xi_i + (x_n-\xi_i)\cos\left(\frac{\mu}{r^2}\right) - y_n\sin\left(\frac{\mu}{r^2}\right)
\end{empheq}

The map represents two point vortices separated by a finite distance $a$ which blink on and off. Here $\xi_i =  \pm 1$ indicate the positions of the vortices $i = 1,2$, so that $a=2$ here, and $\mu$ is the dimensionless flow strength, which is related to strength of the vortex  and the time for which the vortex is operative. The quantity $r = \sqrt{(x-x_i)^2 + (y-y_i)^2}$ is the distance of the point $(x,y)$ from the operating vortex $i$. Fig.~\ref{fig:BV} shows the x-y phase space of the BV map at $\mu = 0.55$ for 25 initial conditions chosen randomly from a uniform distribution on the interval $[-2,2]$. We study the system of coupled blinking vortex maps coupled via the Pecora-Carroll scheme and see the effect of the blinking action on synchronization.

We identify phase synchronization in the system by analyzing the phases via the Hilbert transform \cite{Pikovsky2001,Boccaletti2002} of the time series for both drive and response maps. An easily computable estimate of the phase is the angular variable or the {\it raw}  phase, $\theta = \tan^{-1}\left(\frac{y}{x}\right)$. However, the analytical signal approach based on the Hilbert transform has been found to be more reliable to ascertain phase synchronization in many cases including experimental situations~\cite{Pikovsky2001}. We will, therefore, use this method in our analysis of phase synchronization. We define, $x_d = X_d + iX_d^H$ and  $y_d = Y_d + iY_d^H$ for the drive system and a similar expression for  
the $y_r$ for the response system. The phases are then computed as follows:
\begin{equation}
\hspace{2cm} \phi_d = \tan^{-1}\left(\frac{Y^H_d}{X^H_d}\right),
\phi_r = \tan^{-1}\left(\frac{Y^H_r}{X^H_r}\right)
\end{equation}

\begin{figure}[t!]
\centering
\subfigure[]{\includegraphics[height=7.5cm,width = 8cm]{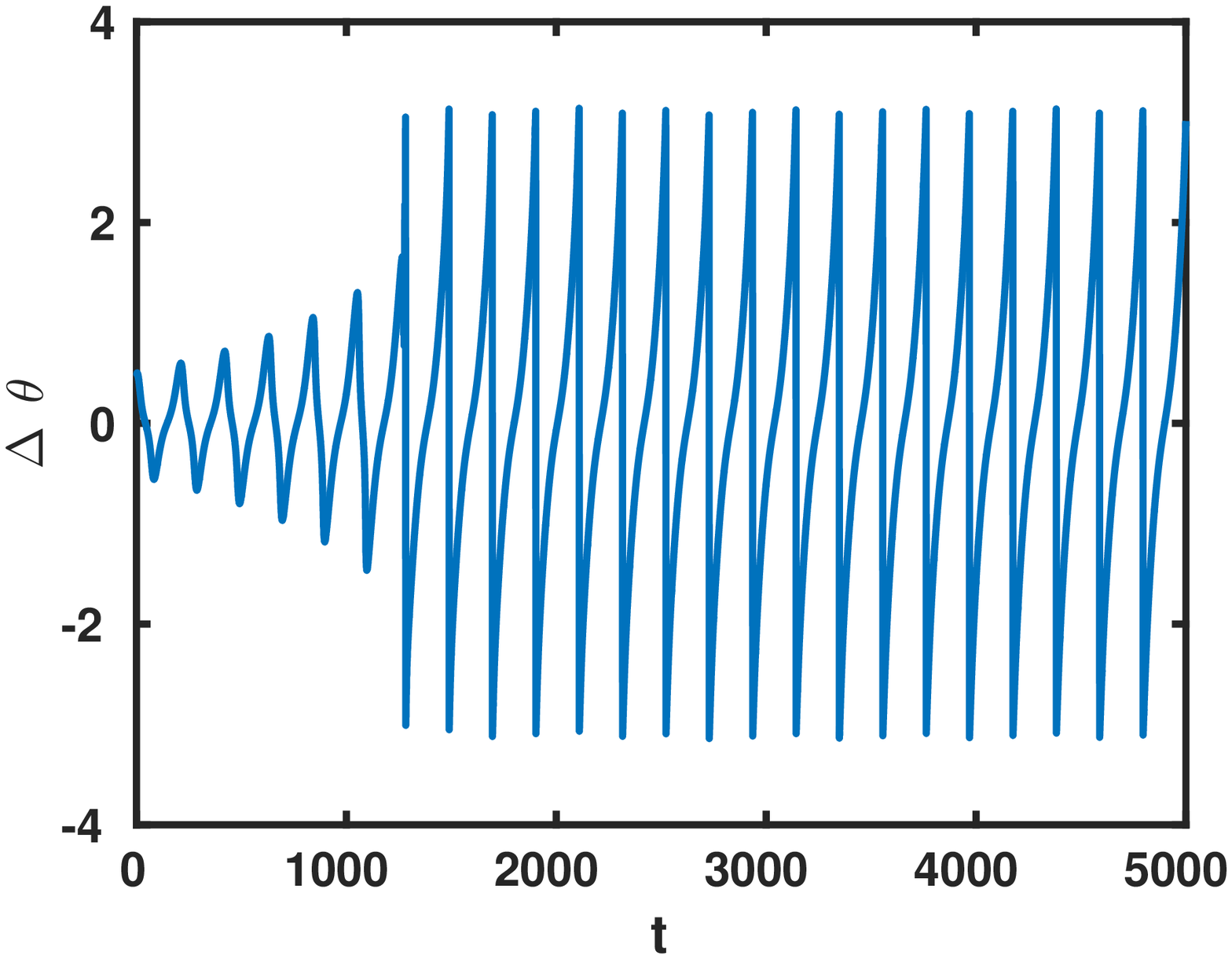}}
\subfigure[]{\includegraphics[height=7.5cm,width = 8cm]{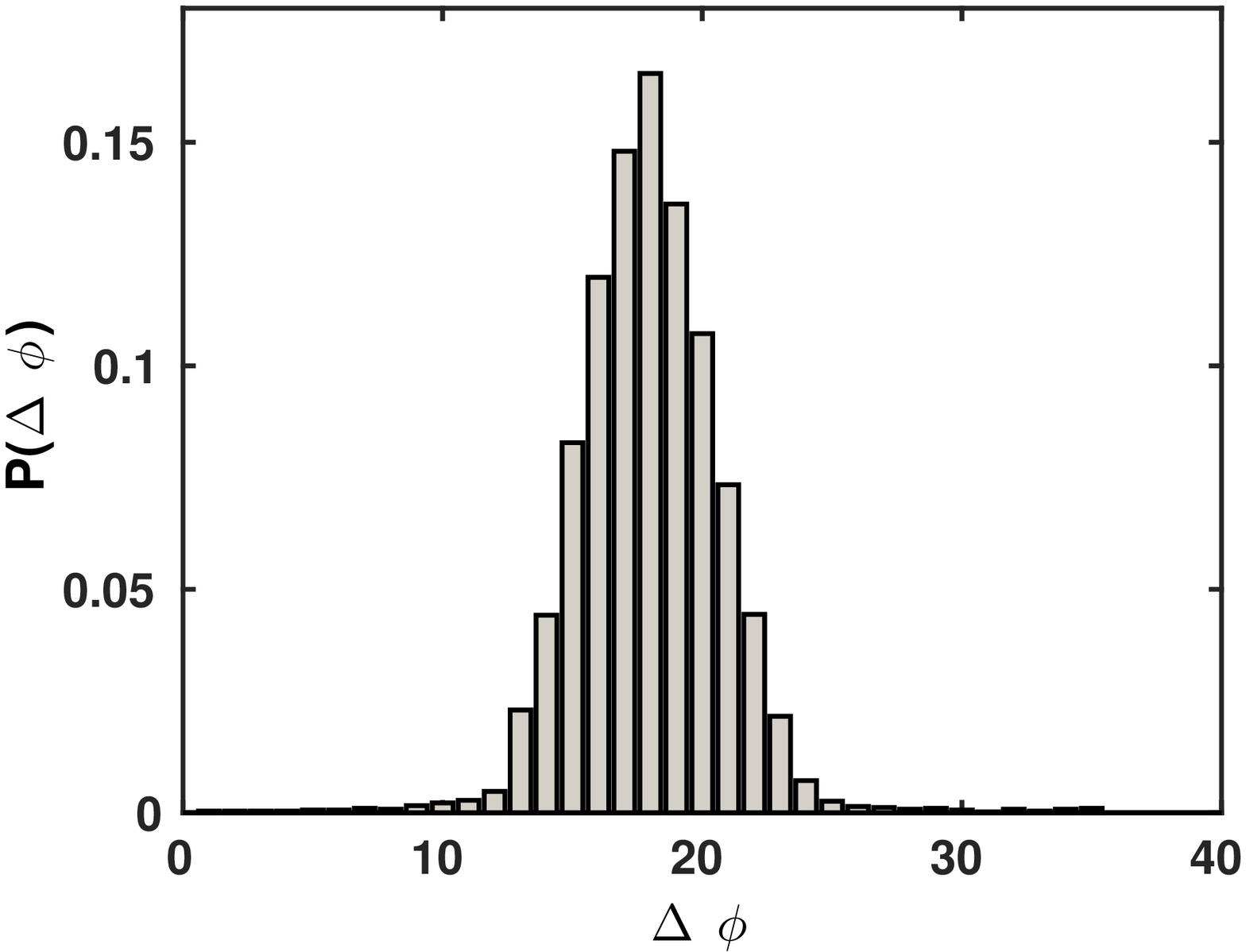}}
\caption{\label{fig:Phase_sync} \footnotesize The coupled BV map system at $\mu = 0.55$  (a) The plot of $\Delta \theta$, the difference in $\theta$; (b) The histogram for  $\Delta\phi$ which shows a peak indicating phase synchronization.  $\Delta \phi$ is calculated using the imaginary part of the Hilbert transform of the original time series (see the text for details). The synchronization index R (defined in the text) takes the value $R=0.92$.}
\end{figure}
For the Pecora-Carroll coupling, $X^H_d = X^H_r$. The signature of synchronization may then be obtained by calculating the difference $\Delta \phi = \phi_d - \phi_r$.  The existence of phase synchronization is confirmed by examining the  histogram for the $\Delta\phi$ distribution which shows a clear peak in case of synchronization. It is further quantified by computing the value of the synchronization index $R = |\frac{1}{N}\sum^N_{j = 1}e^{\Delta \phi_j}|$. This quantity is a measure of the {\it phase coherence}, and takes the value 1, for complete phase synchronization, and the value zero for no synchronization indicating a uniform distribution of phases.

\begin{figure}[t!]
\subfigure[]{\includegraphics[height=7.5cm,width = 8cm]{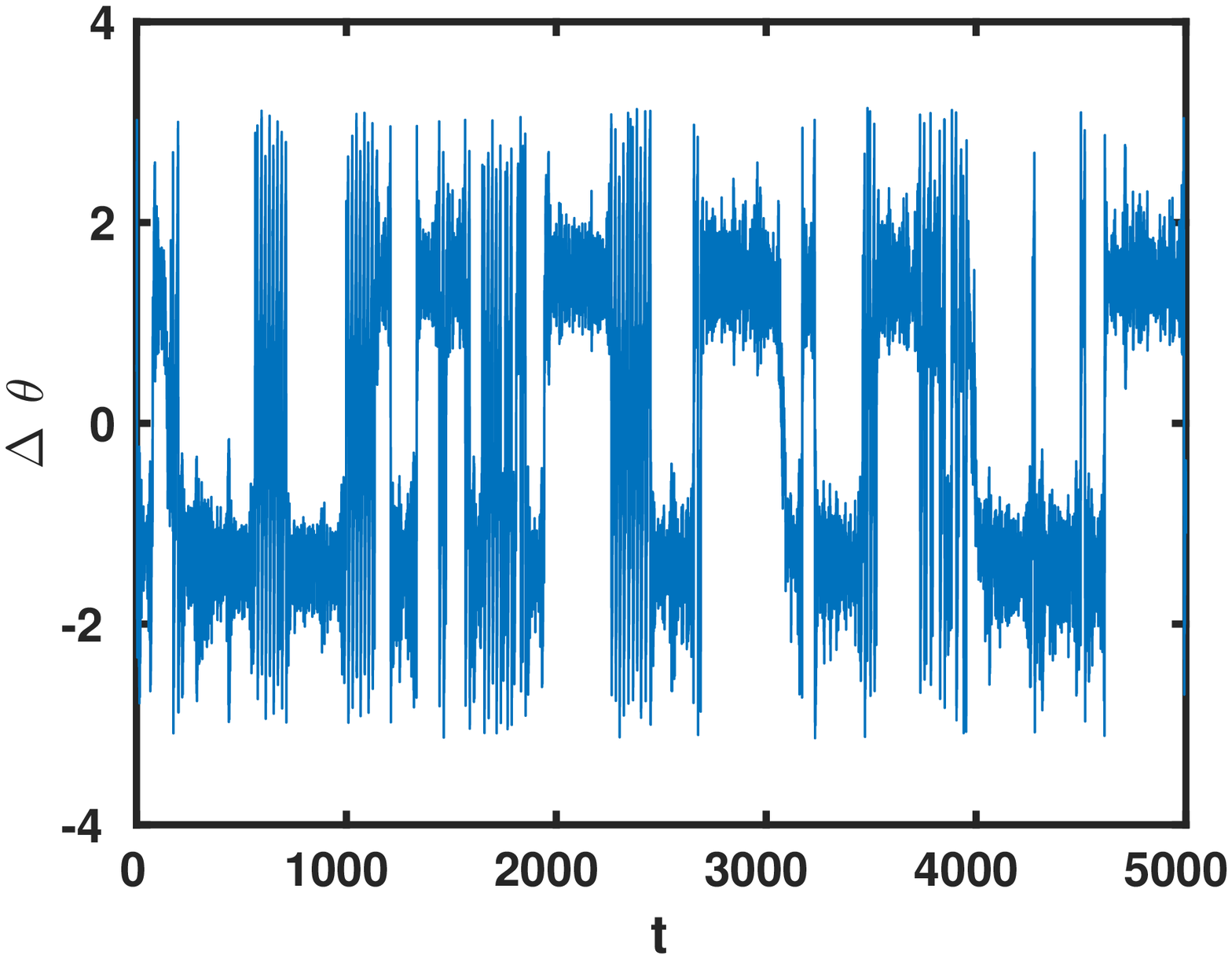}}
\subfigure[]{\includegraphics[height=7.5cm,width = 8cm]{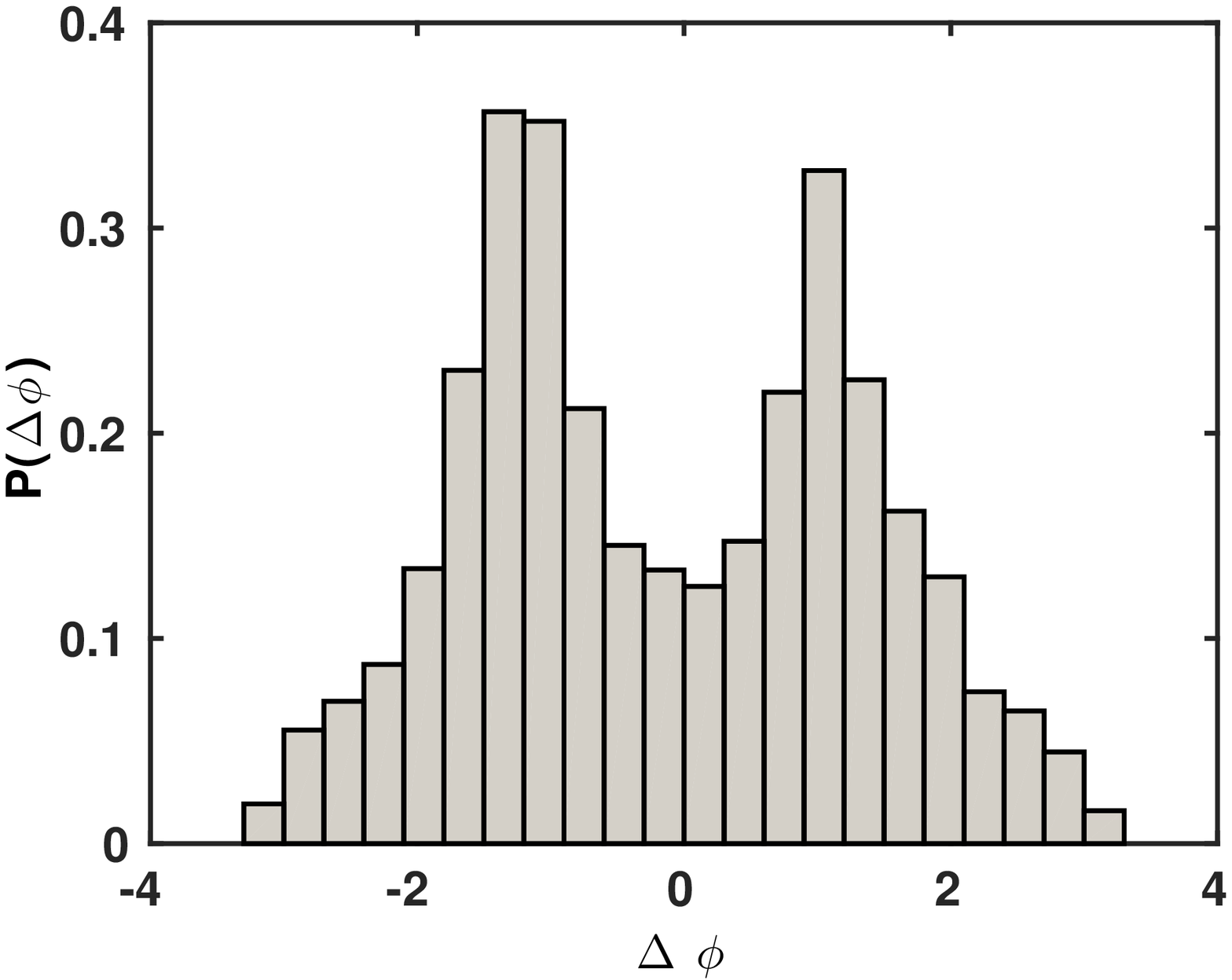}}
\caption{\label{fig:Phase_slip} \footnotesize Phase slips in the coupled BV map system at $\mu = 0.55$  (a) The plot of the time-series of the phase difference $\Delta \theta$. Phase slips are clearly seen; (b) The  histogram for $\Delta\phi$ shows {\it two} peaks indicating {\it almost} phase synchronized states of positive and negative phases. $\Delta \phi$ is calculated using the imaginary part of the Hilbert transform of the original time series (see the text for details). The synchronization index R (defined in the text) for positive phases is $0.78$ and that of negative phases is $0.77$. Our empirical observations suggest that this typically remains greater than $0.7$.}
\end{figure}

In the case of the Pecora-Carroll  coupled BV maps, we observe the phenomenon of phase synchronization. Fig.~\ref{fig:Phase_sync}(a) plots a  time series for the difference in the {\it raw} phases i.e. $\Delta \theta = (\theta_d - \theta_r) $ for the drive and response maps at $\mu = 0.55$ where   the raw phase is defined as $\theta = \tan^{-1}\left(\frac{y}{x}\right)$. The corresponding $\Delta\phi$-histogram in Fig.~\ref{fig:Phase_sync}(b) shows a clear peak. The value of the  synchronization index is found to be $R = 0.92$. 

Interestingly,  for certain initial conditions which belong to the interval $[-1,1]$, we observe the  phenomenon  of phase slips, see Fig~\ref{fig:Phase_slip}(a).  The $\Delta\phi$-histogram for an instance of phase slip typically shows two peaks (see Fig~\ref{fig:Phase_slip}(b)). This indicates that the system continuously switches between two {\it almost} phase synchronized states as $R$ for positive and negative phases are usually less than $0.8$. The phase slips seen here appear to be the effect of the alternatively blinking action of the vortices on the trajectories evolving from the initial conditions that lie between the vortices. For initial conditions outside this range, such slips are not found indicating that their trajectories experience an overall effect of the stirring only. We see almost regular oscillations of the phase difference $\Delta \theta$  for such initial conditions unlike the abrupt jumps in the slip regime.  No phase slips were seen in the case of the coupled standard maps studied in the earlier section; due to the absence of any analogue of the blinking action. 

\section{Conclusion}
\label{sec:conclusion}
To summarize, we explored different aspects of phase synchronization observed in coupled area -preserving maps, coupled by the Pecora-Carroll method.  For coupled standard maps, the distribution of synchronized times depends crucially both on the value of the non-linearity parameter $K$ as well as on the mixed nature of the phase space. Distributions of synchronized times show long tailed behavior with accompanying power law for initial conditions distributed uniformly in the phase space. The introduction of a coherent structure in the system drastically alters the distribution of synchronization times, which crosses over to exponential behavior, and also drastically enhances the size of the basin of synchronization. Thus, the introduction of a coherent structure can be an effective strategy in cases where short synchronization times are desirable for area-preserving systems. 

The study of the blinking vortex map illustrates the phenomena of phase slips, which are not seen in the case of the standard map. This is a consequence of the switching nature of the BV map, which does not exist in the case of the standard map. We also note that initial conditions which give rise to trajectories which do not wander in the vicinity of the switching action do not show phase slips. This again is due to the inhomogeneity in the phase space, which is divided into two regions, one of these being the region which feels the switching action, and the other of which does not. Thus, the behavior of the BV map also illustrates the manner in which the inhomogeneity in the system gives rise to different kinds of synchronization behavior. The BV map is also a model of a continuously stirred reactor, and is an important system from the point of view of applications. For instance, the existence of both these two phenomena -- phase slips and phase synchronization, has been seen in an  experimental study \cite{Pita2010} of a simple laboratory system where two chaotically rotating annuli thermally coupled in a drive-response configuration (similar to  the Pecora-Carroll scheme). These experimental studies were carried out to model phenomena which have been seen in atmospheric systems in the presence of teleconnections. It would be interesting to examine whether the implications of our study, viz. the importance of initial conditions, and the effect of switching have relevance for these experimental contexts.  Moreover, such phase slips have also been known to be characteristic of coupled systems entering a synchronized Arnold's tongue~\cite{Boccaletti2002}. We hope to pursue these connections elsewhere.




\end{document}